\documentclass[twocolumn,nofootinbib,showpacs]{revtex4}
\usepackage{nccmath}
\usepackage[T1]{fontenc} 
\usepackage{ae} 
\usepackage[ansinew]{inputenc}
\usepackage{tensor}
\usepackage{graphicx}
\usepackage{caption}
\usepackage{subcaption}
\usepackage{placeins}
\input{epsf}

\def\to{\rightarrow}

   \def\de{\delta}
   
\def\th{\theta}

\def\lsim{\mathrel{\rlap{\lower4pt\hbox{\hskip1pt$\sim$}}
\raise1pt\hbox{$<$}}}
\def\gsim{\mathrel{\rlap{\lower4pt\hbox{\hskip1pt$\sim$}}
\raise1pt\hbox{$>$}}} \def\sqr#1#2{{\vcenter{\vbox{\hrule height.#2pt
\hbox{\vrule width.#2pt height#1pt \kern#1pt \vrule width.#2pt} \hrule
height.#2pt}}}}

\def\beq{\begin{equation}} \def\eeq{\end{equation}}
\def\beqa{\begin{eqnarray}} \def\eeqa{\end{eqnarray}}

\let\oldfrac\frac
\let\frac\dfrac

\begin{document}

\title{Astrophysical constraints on the bumblebee model}

\author{Gon\c{c}alo Guiomar}
\affiliation{Departamento de F\'{\i}sica, Instituto Superior T\'ecnico, Universidade de Lisboa,\\Avenida Rovisco Pais 1, 1049-001, Lisboa, Portugal.}
\email{goncalo.guiomar@ist.utl.pt}

\author{Jorge P\'aramos}
\affiliation{Departamento de F\'isica e Astronomia and Centro de F\'isica do Porto,\\Faculdade de Ci\^encias, Universidade do Porto, \\Rua do Campo Alegre 687, 4169-007 Porto, Portugal.}
\email{jorge.paramos@fc.up.pt}

\date{\today}

\begin{abstract}

In this work the bumblebee model for spontaneous Lorentz symmetry breaking is considered in the context of spherically symmetric astrophysical bodies. A discussion of the modified equations of motion is presented and constraints on the parameters of the model are perturbatively obtained.
\end{abstract}


\maketitle 


\section{Introduction}

One of the basic assumptions of general relativity is that of Lorentz invariance, a fundamental symmetry which to the present day has been verified to a very high precision. However, the possibility of this symmetry being broken is an ongoing topic of debate \cite{Liberati:2013xla,Mattingly:2005re}: a relevant branch of this discussion is centered on the consequences that Lorentz symmetry breaking (LSB) would have in gravitation. The exploration of these consequences can be made through Kosteleck\'y's standard model extension \cite{Kostelecky:2003fs} which, as the name implies, extends the scope of the standard model by adding a gravitational sector along with Lorentz-violating terms \cite{Bluhm:2008rf}.

The spontaneous Lorentz-breaking mechanism is similar to the Higgs mechanism, where the system spontaneously collapses onto the referred vacuum expectation value (VEV), achieving thus a particular four-vector orientation and creating preferred frame effects as a consequence. 

This can modeled through the introduction of a vector field dynamically driven by a potential which acquires a nonvanishing VEV: notice that, in the context of the pioneering work developed in Refs. \cite{Kostelecky:2003fs,Bluhm:2008rf}, this vector field is not simply a classical addition to the matter content of the model, but instead is assumed to arise dynamically from the LSB terms in the standard model extension, the underlying quantum field theory.

The focus of this work will be mostly on the LSB in the context of gravity, using a subset of the so called Einstein-aether theories\cite{Zlosnik2007,Tartaglia2007} - the bumblebee models - where a vector field with a nonvanishing VEV is added to the Einstein-Hilbert action of the system. The experimental constraints obtained in both high-energy cosmic rays \cite{Galaverni:2007tq} and gravitational experiments (the latter through the post-Newtonian formalism \cite{Will:2005va}) show very little lenience in allowing the breaking of this symmetry \cite{Bertolami:2012ta}. The study of the implications of LSB in gravity through Einstein-aether theories has only recently been explored \cite{ArmendarizPicon:2009ai,Jacobson:2008aj}, and the impact of the bumblebee model on Solar System dynamics was assessed in Ref. \cite{Bertolami:2005bh}.

The purpose of this work is to apply this model to the study of astrophysical bodies such as stars, in order to obtain a lower bond on the parameters of the model.

The action for the bumblebee model is given by

\begin{eqnarray}\label{action}
S=\int d^4 x \bigg[ && {\cal L} +  \frac{1}{16\pi G} (R+\xi  B^\mu B^\nu R_{\mu\nu}) - \\ \nonumber && \frac{1}{4} B^{\mu\nu} B_{\mu\nu} - V(B^\mu B_\mu\pm b^2) \bigg],
\end{eqnarray}

\noindent where ${\cal L} $ is the Lagrangian density of matter, $B_{\mu\nu}$ is the field strength,

\begin{equation}
B_{\mu\nu} = \nabla_\mu B_\nu - \nabla_\nu B_\mu,
\end{equation}

\noindent and $\xi$ the coupling constant between curvature and the Bumblebee field; the potential $V$ has a nonvanishing VEV $b \neq 0$ signalling the spontaneous Lorentz symmetry breaking.

\section{The Model}

The variation of Eq. (\ref{action}) with respect to the metric yields the modified equations of motion \cite{Kostelecky:2003fs},

\begin{equation}\label{modfieldeqs}
\begin{split}
R_{\mu\nu} - {1 \over 2} R g_{\mu\nu} = 8\pi G \left( \tensor[]{T}{^M}{_{\mu\nu}} + \tensor[]{T}{^B}{_{\mu\nu}} \right), 
\end{split}
\end{equation}

\noindent where $\tensor[]{T}{^M}{_{\mu\nu}}$ is the matter stress-energy tensor and  $\tensor[]{T}{^B}{_{\mu\nu}}$ is the bumblebee stress-energy tensor, defined as

\begin{equation}
\begin{split}
\tensor[]{T}{^B}{_{\mu\nu}} \equiv -B_{\mu\alpha} \tensor[]{B}{^\alpha}_{\nu}-\frac{1}{4}B_{\alpha\beta}B^{\alpha\beta}g_{\mu\nu} -Vg_{\mu\nu}+ \\ 2V'B_\mu B_\nu +
{\xi \over 8\pi G} \bigg[ \frac{1}{2} B^\alpha B^\beta R_{\alpha\beta} g_{\mu\nu} - B_\mu B^\alpha R_{\alpha\nu} \\ +\frac{1}{2}\nabla_\alpha \nabla_\mu (B^\alpha B_\nu)+ \frac{1}{2}\nabla_\alpha \nabla_\nu (B^\alpha B_\mu) \\ -\frac{1}{2} \nabla^2 (B_\mu B_\nu)- \frac{1}{2}g_{\mu\nu}\nabla_\alpha  \nabla_\beta (B^\alpha B^\beta) \bigg].
\end{split}
\end{equation}

\noindent No separate conservation laws are assumed for matter and the bumblebee vector field. The covariant (non)conservation law (which is not used) can be obtained directly from the Bianchi identities $\nabla^\mu G_{\mu\nu}$ applied to both sides of the modified field equations (\ref{modfieldeqs}): this leads to $\nabla^\mu T^M_{\mu\nu} = - \nabla^\mu T^B_{\mu\nu}  \neq 0 $, which may be interpreted as an energy transfer between the bumblebee and matter.

The equations for the bumblebee field are

\begin{equation}
\label{beq1}\nabla_\mu B^{\mu\nu}=2V' B^\nu -\frac{\xi}{8\pi G}B_\mu R^{\mu\nu},
\end{equation}

\noindent where a prime represents differentiation with respect to the argument. 

A potential of the form 

\begin{equation}
V=A (B_{\mu}B^{\mu}\pm b^2)^n,
\end{equation}

\noindent is assumed, so  that the Bumblebee field (\ref{beq1}) becomes

\begin{equation}
\label{beq2}
B^{\mu}[ 16\pi G V' g_{\mu\nu}-\xi R_{\mu\nu}]=0.
\end{equation}

\section{Static, spherically symmetric scenario}

Given that the relevant quantities such as the density, pressure and scalar curvature inside a spherical symmetric body such as the Sun have a strong radial variation when compared with very slow temporal changes, one assumes that the bumblebee field is given by

\begin{equation}
B^\mu = (0,B(r),0,0).
\end{equation}

\noindent Accordingly, one resorts to the static Birkhoff metric,

\begin{equation}
g_{\mu\nu} = diag \left[ -e^{2\nu(r)},\left( 1 - \oldfrac{2 G m}{r}\right )^{-1},r^2,r^2 \sin^2  \theta \right],
\end{equation}

\noindent where $m(r)$ is the mass profile as a function of the radial coordinate, and we assume that the potential takes a quadratic form, for simplicity,

\begin{equation}
V=A (B_{\mu}B^{\mu}-b^2)^2,
\end{equation}  

\noindent with the adopted sign reflecting the spacelike nature of the bumblebee field.

For the radial case $\mu=r$, the Ricci tensor is given by,

\begin{equation}
\begin{split}
R_{rr}= {G (m'r -m) ( 2 + r \nu') \over r^2(r-2Gm)} -(\nu')^2 - \nu''.
\end{split}
\end{equation}

\noindent The only nonvanishing component of Eq. (\ref{beq2}) is for $\mu = r$, yielding 

\begin{eqnarray}\label{Bumblebeeeq0}
&& 16\pi G V' g_{rr} - \xi R_{rr} = 0 \to \\ \nonumber 2A(g_{rr}B^2-b^2)&=& \frac{\xi}{16\pi G r^3} \bigg[G r^2 [m'\nu' +2m(\nu''+\nu'^2)] + \\ \nonumber &&
G r(2m'-m\nu')-2m G-r^3(\nu''+\nu')\bigg].
\end{eqnarray}

This gives us, after some algebraic manipulation

\begin{equation}\label{Bumblebeeeq}
\begin{split}
B^2=\left(1 - \frac{2G m}{r}\right) \left[ b^2+\frac{\xi}{32 \pi G A r^3}( G r^2[m'\nu'\right. \\ \left.+2m(\nu''+\nu'^2)] \vphantom{\frac{1}{g_{rr}}}
+G r(2m'-m\nu') - \right. \\ \left. 2 Gm -r^3[\nu''+\nu'])\vphantom{\frac{\xi}{32 \pi G A r^3}} \right].
\end{split}
\end{equation}

In order to obtain the pressure and density equations, we resort to the trace-reversed field equations, given by,

\begin{equation}
\label{MEE}E_{\mu\nu} \equiv R_{\mu\nu}-8\pi G \left[ T^M_{\mu\nu} + T^B_{\mu\nu} -\frac{1}{2}g_{\mu\nu}(T^M +T^B) \right ]=0,
\end{equation}

\noindent where $T^M$ and $T^B$ are traces of the stress-energy tensors for normal matter and the bumblebee field, respectively.

The stress-energy tensor for normal matter is given by the perfect fluid form,

\begin{equation}
\begin{split}
T^M_{\mu\nu}=(\rho + p) u_\mu u_\nu + p g_{\mu\nu} ,
\end{split}
\end{equation}

\noindent where $u_\mu$ is the four-velocity; in the static scenario and given that $u_\mu u^\mu = -1$, we have $u_\mu = (e^{\nu(r)}, \vec{0}) $, so that

\begin{equation}
T^M_{\mu\nu}= \text{diag} \left( e^{2 \nu} \rho, \oldfrac{p}{1-\oldfrac{2 G m}{r}},p r^2 , p r^2 \sin ^2(\theta )\right),
\end{equation}

\noindent with trace $T = 3p-\rho$.

Using 
\begin{equation}
g^{tt}E_{tt}-g^{rr}E_{rr}=g^{\theta\theta}E_{\theta\theta}=0,
\end{equation}

\noindent one may derive the equations that will allow us to obtain $p(r)$, $\rho(r)$ and $\nu(r)$. Without the bumblebee field, these quantities (denoted with the subscript $_0$) are given by 

\begin{eqnarray}\label{unperturbed}
p_0(r)=\frac{ r (r-2 G m_0) \nu_0' - G m_0}{4 \pi  G r^3}, \\
\rho_0(r)=\frac{m_0'}{4\pi r^2} ,\\
\nu_0'(r)= G\frac{m_0 + 4\pi p_0  r^3}{r(r-2 G m_0)},
\end{eqnarray}

\noindent which, along with a state equation that relates $p_0$ and $\rho_0$, yields a closed set of four differential equations with four unknowns.

In the presence of the bumblebee field, one must also include the related field (\ref{beq2}); solving Eq. (\ref{MEE}), the pressure and density are then given by

\begin{equation}\label{pressureeq}
\begin{split}
p(r)=\frac{1}{8\pi G r^4}\left [ r\xi B (r-2G m)^2 B'(2+r\nu') \right. \\ \left.
 +r( 8 \pi G V r^3 - 2G m+2r(r-2G m)\nu')- \right. \\ \left. B^2(r-2G m)(-2\xi G m(-1+r(\nu'(-2+r\nu') \right. \\ \left.+r\nu''))+
r[ 16 \pi GV'r^2 \right. \\ \left. -\xi +r\xi (\nu'(-2+r\nu')+r\nu'')])\right],
\end{split}
\end{equation}

\noindent and

\begin{equation}\label{densityeq}
\begin{split}
\rho(r)= \frac{1}{8\pi G r^4}[-r^2(8 \pi G V r^2 + \xi(r-2G m)^2B'^2\\-2G m')+ r\xi B(r-2G m)(B'(-4r+3G m\\+5G r m') -r(r-2G m)B'')+ \xi B^2(3 G^2 m^2\\-2G^2 r m(3m'+rm'') \\+r^2[-1+G(m'(4-G m')+rm'')]) ].
\end{split}.
\end{equation}

Although we have a complete set of equations that describe the behavior of our system, the solution of that set of equations implies very intensive numerical computations. Instead, since the stellar structure of the Sun is known to be well described by general relativity, one shall adopt a perturbative approach.

\subsection{Lane-Emden solution}

In order to obtain the perturbations to the pressure and density arising from the effect of the Bumblebee field, one first describes how these quantities are obtained in the standard scenario of General Relativity. To do so, one first writes the Tolman-Oppenheimer-Volkov equation, derivable from Eq. (\ref{unperturbed}),

\begin{equation}
{dp \over dr} = -G (\rho+p){ m + 4\pi p r^3 \over r ( r - 2G m) }.
\end{equation}

\noindent In the Newtonian regime (invalid for relativistic neutron stars but a good approximation for main sequence stars such as the Sun), specified by the following conditions,

\begin{equation}
p(r) \ll \rho(r)~~,~~4\pi p(r) r^3 \ll m(r)~~,~~2Gm(r) \ll r,
\end{equation}

\noindent the above equation may be approximated by the hydrostatic equilibrium condition,
\begin{equation}
{dp \over dr} = - { G m \rho \over r^2}.
\end{equation}

The first model for the internal structure of the Sun was put forward by Eddington, assuming that solar matter is described by a polytrope equation of state (EOS),

\begin{equation}\label{EOS}
p = K \rho^{1+1/n}~~,~~K = {p_c \over \rho_c^{1+1/n}},
\end{equation}

\noindent with polytropic index $n=3$, and where $\rho_c$ and $p_c$ are the central density and pressure, respectively. Although this is a crude model, surpassed by state of the art numerical models of the several layers and processes occurring inside the Sun, it is well suited for analytical studies and serves the purpose of our study: obtaining bounds for the parameters of the model under scrutiny compatible with a perturbative impact on the interior structure of our star (as shown for scalar field-  \cite{PhysRevD.71.023521} and ungravity-inspired models \cite{Bertolami:2009jq}).

With the above EOS, the hydrostatic equilibrium condition gives rise to the Lane-Emden (LE) equation,
\begin{equation}
\label{LEeq}
\frac{1}{\chi^2}\frac{d}{d\chi}\left ( \chi^2\frac{d\theta}{d\chi} \right ) = -\theta^n .
\end{equation}

\noindent  Here $\theta(\chi)$ is a dimensionless function which gives us the density and pressure profiles of the system through	
\begin{equation}\label{LEdefs}
\rho = \rho_c \theta^n(\chi)~~,~~p = p_c \theta^{n+1}(\chi),
\end{equation}

\noindent where $\chi = r / r_n$ is a dimensionless coordinate, and

\begin{equation}
r_n^2 = \frac{(n+1)p_c}{4\pi G \rho_c^2}.
\end{equation}

\noindent The boundary of the spherical body occurs at $\chi = \chi_f$, so that $\theta(\chi_f)=p(\chi_f)=\rho(\chi_f)=0$; this is related with its physical radius $R$ by $\chi_f = R/r_n$.

\begin{figure}[htp]
\centering
\includegraphics[width=\linewidth]{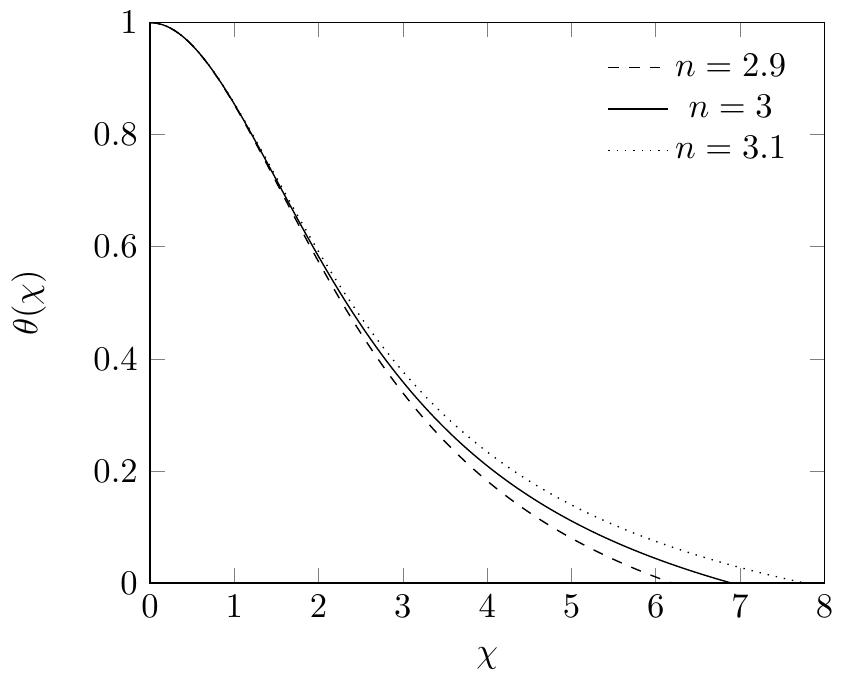}
\caption{Lane-Emden solution for three polytropic indices.}
\label{figLE}
\end{figure}

The solution $\th(\chi)$ of the LE equation (\ref{LEeq}) for $n\sim 3$ is plotted in Fig. \ref{figLE}, showing {\it e.g.} that $\chi_f \approx 6.9$ for $n=3$. Notice that this quantity depends solely on the polytropic index $n$, as does $\th$: the physical quantities $\rho_c$, $p_c $ and $R$ affect only the value of $r_n$. This reflects the homology symmetry of the LE equation, so that all stars with the same polytropic index $n$ share a common density (and pressure or temperature) profile, scaled only by its central value.

One may read the (unperturbed) mass of the spherical body from the relation $\rho_0(r)=m'/4\pi r^2$, obtaining

\begin{equation}
\begin{split}
m(r)= 4\pi \int_0^r \rho(r)r^2 dr= \\ 4\pi r_n^3 \rho_c \int_0^{\chi} \chi^2 \theta^n d\chi =  - 4\pi \rho_c r_n^3 \chi^2 \theta',
\end{split}
\end{equation}

\noindent where the LE equation (\ref{LEeq}) was used; the total mass of the star is given by replacing $\xi = \xi_f$, $M = - 4\pi \rho_c r_n^3 \chi_f^2 \theta'(\chi_f)$.

\section{Perturbative Effect of the Bumblebee Field}

The linearization of this system of equations (\ref{Bumblebeeeq},\ref{pressureeq},\ref{densityeq}) brings with it a very large complexity that makes it very computationally demanding to solve. With that in mind, we consider the perturbation to be of zeroth order, {\it i.e.} we replace the quantities on the {\it r.h.s.} of the equations mentioned above by the unperturbed expressions for $ m_0(r)$ and $\nu_0(r)$ and the bumblebee field. Regarding the latter, it is more straightforward to resort instead to Eq. (\ref{Bumblebeeeq0}), since at zeroth order one has 

\begin{equation}
R_{rr}=8\pi G\left( T^M_{rr}-\frac{1}{2}g_{rr}T^M \right) = 4\pi G (\rho_0-p_0 )g_{rr} ,
\end{equation}

\noindent which leads to 

\begin{equation}
B^2(r)= \left(1 - {2Gm\over r}\right) \left( b^2 + {\rho_0-p_0 \over 8A}g_{rr}\right) .
\end{equation}

\noindent Since the unperturbed solutions $\rho_0(\chi) $ and $p_0(\chi)$ vanish at the boundary of the spherical body, the above shows that the bumblebee field collapses onto its VEV as it crosses to its outer solution (where $T^M_{\mu\nu} = 0$), $B_\mu B^\mu = b^2$. This is consistent with the approach followed in Ref. \cite{Bertolami:2005bh}, where the latter condition was also assumed.

Following the above procedure, the expressions for the pressure and density may be obtained, yielding
\begin{widetext}
\begin{equation}
\begin{split}
p(r)=p_0 + [\xi (p-\rho)]^2 + {\xi^2 (\rho'-p')\over 2A G \pi r}  \left( 1 - {2Gm\over r}\right)^3 (2+\nu'r)  + \\ {\xi \over A \pi r^3} \left( 1 - {2Gm\over r}\right)^2 [8A b^2 + \xi (\rho-p)] \times \bigg[ (2 + \nu' r) (m-m' r) + \\ {r \over G }  \left( 1 - {2Gm\over r}\right) \left( 1 + r (4 G \pi p r + 2 \nu' - r [ 4 G \pi \rho + (\nu')^2 + \nu'' ] ) - {2 G m \over r} [ 1 + r (\nu' [ 2 - \nu' r ] - r \nu'' ) ] \right) \bigg],
\end{split}
\end{equation}

\begin{equation}
\begin{split}
\rho(r)= \rho_0 - \left[{\xi \over 8} (\rho - p)\right]^2 + {\xi^2 \over 128 A G \pi r} \left( 1 - {2Gm \over r} \right)^2 \left[ \left( 4 + G \left[{ m\over r} - 9 m' \right] \right) ( p' - \rho' ) + \left(1 - {2Gm\over r}\right)^3 (p'' - \rho'') r\right]+ \\ {\xi \over 64 A G \pi r^2} \left( 1 - {2Gm \over r} \right) \left[ 8 A b^2 + \xi ( \rho-p) \right] \left[ 2\left({Gm \over r}\right)^2 +  6 G m' (1 - Gm') + 2 G m'' r - 1 - {2Gm \over r} (1+2Gm'' r)\right],
\end{split}
\end{equation}
\end{widetext}

The advantage of considering the admittedly simplistic model provided by the polytropic EOS (\ref{EOS}) lies in the possibility of rewriting the rather convoluted expressions above in terms of the LE solution $\th(\chi)$ only. For this, we now introduce the dimensionless parameters

\begin{equation}
 \alpha \equiv \frac{\xi^2}{R^2 G}, \, \beta \equiv \frac{\xi^3 b^2}{R^2 G},\,\gamma \equiv \frac{R_s}{R},
\end{equation}

\noindent where $R_s \equiv 2 G M$ is the Schwarzschild radius of the star, together with the form factor

\begin{equation}
\phi  \equiv \frac{3M}{4 \pi \rho_c R^3},
\end{equation}

\noindent and the EOS parameter $\omega_c \equiv p_c/\rho_c $.

Using the relations (\ref{LEdefs}) and the expression for $\nu_0'(r)$ from equation (\ref{unperturbed}), one obtains

\begin{equation}
\nu'_0(\chi) =\frac{3\gamma(\chi \omega_c\theta^{1+n}-\theta')}{2\phi\chi_f^2+6\gamma\chi\theta'},
\end{equation}

\noindent and the form factor becomes $ \phi = -3 \theta'(\chi_f)/ \chi_f$ --- again displaying the homology invariance of the LE Eq. (\ref{LEeq}).

We may rewrite the above expressions for the pressure and density in terms of the LE solution $\th(\chi)$ and its derivatives only, obtaining a more manageable form: separating the contributions to the pressure and density arising from the nonvanishing VEV $b$ and the potential strength $ A$ as

\begin{eqnarray}\label{contributions}
p(\chi)=p_0(\chi) + p_b(\chi)+p_{V}(\chi)+\delta(\chi), \\ \nonumber
\rho(\chi)=p_0(\chi) + \rho_b(\chi)+\rho_{V}(\chi)-\delta(\chi),
\end{eqnarray}

\noindent we have

\begin{widetext}
\begin{equation}
\begin{split}
p_b(\chi)=\frac{\beta}{16\pi\phi^3\xi^2\chi_f^4\chi}(\phi\chi_f^2+3\gamma\chi\theta')^2 \times \\ [3\gamma\chi^2\theta^n(-1+\omega_c \theta) -2(-1+\chi[\nu'(-2+\chi\nu')+\chi\nu''])(\phi\chi_f^2+3\gamma\chi\theta')+3\gamma\chi(2+\chi\nu')(\theta'+\chi\theta'')],
\end{split}
\end{equation}

\begin{equation}\label{rhob}
\begin{split}
\rho_{b}(\chi)=\frac{\beta}{16\pi\phi^3\xi^2\chi_f^4\chi}(\phi\chi_f^2+3\gamma\chi\theta')\times \\ [2\phi^2\chi_f^4+3\gamma\chi(45\gamma\chi\theta'^2+\chi[14\phi\chi_f^2\theta'' +9\gamma\chi^2\theta''^2+2\phi\chi_f^2\chi\theta'''] +2\theta'[7\phi\chi_f^2+3\gamma\chi^2(10\theta''+\chi\theta''')])],
\end{split}
\end{equation}

\begin{equation}
\begin{split}
p_{V}(\chi)=\frac{3\gamma\alpha^2\theta^n}{1024\pi^2 A\phi^4\xi^2\chi_f^4\chi^2} (\phi\chi_f^2+3\gamma\chi\theta')^2 \times \\ \bigg[(\omega_c - 1)(3\gamma\chi^2\theta^n[1-\omega_c\theta]+2[-1+\chi(\nu'[ \chi\nu' - 2]+\chi\nu')](\phi\chi_f^2+3\gamma\chi\theta'))-\\ \frac{\chi}{\theta}(2+\chi\nu')(\theta'[\theta([1+n]\phi\chi_f^2\omega_c +3\gamma[\omega_c\theta-1])+3\gamma\chi([1+n]\omega_c\theta - n)\theta' -n\phi\chi_f^2 ]+3\gamma\chi\theta[\omega_c\theta - 1]\theta'')\bigg],
\end{split}
\end{equation}

\begin{equation}
\begin{split}
\rho_{V}(\chi)=\frac{3\gamma\alpha^2\theta^n}{2048\pi^2A\phi^4\xi^2\chi_f^4\chi^2}(\phi\chi_f^2+3\gamma\chi\theta') \times \\ \bigg[\frac{\chi}{\theta^2}(\phi\chi_f^2+3\gamma\chi\theta')(2[1-n]n\chi\theta'^2[\phi\chi_f^2+3\gamma\chi\theta'] + n\theta[\theta'(\chi\theta'[-51\gamma+2(1+n) \phi\chi_f^2\omega_c+6\gamma(1+n)\chi\omega_c\theta']-8\phi\chi_f^2)- \\ \chi(2\phi\chi_f^2+33\gamma\chi\theta')\theta'']+[1+n]\omega_c\theta^2[\theta'(8\phi\chi_f^2+51\gamma\chi\theta')+\chi(2\phi\chi_f^2+33\gamma\chi\theta')\theta''])\\-2(2\phi^2\chi_f^4+3\gamma\chi[45\gamma\chi\theta'^2+\chi(14\phi\chi_f^2\theta''+9\gamma\chi^2\theta''^2+2\phi\chi_f^2\chi\theta''')+2\theta'(7\phi\chi_f^2+3\gamma\chi^2[10\theta''+\chi\theta'''])])+ \\ 2\omega_c\theta(2\phi^2\chi_f^4+3\gamma+\chi[45\gamma\chi\theta'^2+\chi(14\phi\chi_f^2\theta''+9\gamma\chi^2\theta''^2 +2\phi\chi_f^2\chi\theta''')+2\theta'(7\phi\chi_f^2+3\gamma\chi^2[10\theta''+\chi\theta'''])])\bigg],
\end{split}
\end{equation}

\end{widetext}

\noindent and

\begin{equation}\label{delta}
\de(\chi)=\frac{9\gamma^2\alpha^2\theta^{2n}( \omega_c\theta - 1)^2}{4096 \pi^2\phi^2\xi^2}.
\end{equation}

\noindent The latter appears in both the pressure and density perturbations and, as shall be shown, has a negligible impact when compared with the remaining contributions. 

\begin{figure*}[!htp]
        \begin{subfigure}[b]{0.3\textwidth}
                \hspace*{-4.5cm}\includegraphics[scale=1.5]{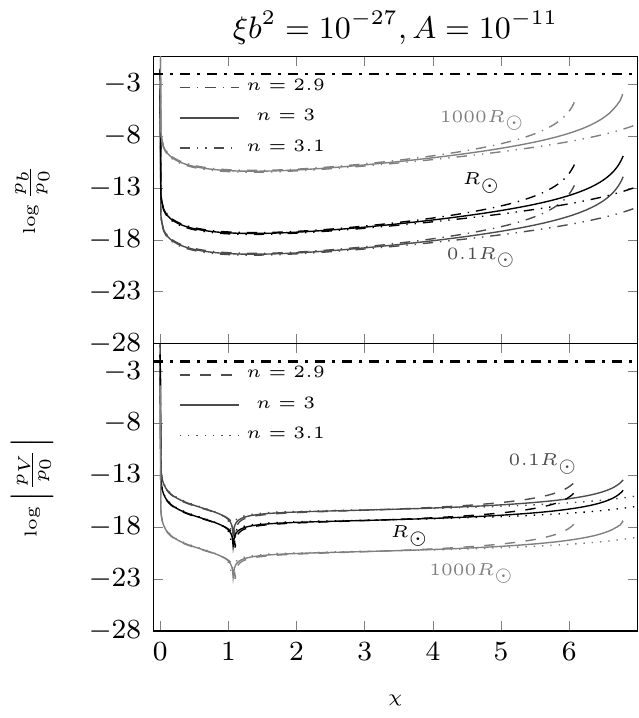}
        \end{subfigure}%
        \quad
        \begin{subfigure}[b]{0.3\textwidth}
                \hspace*{-0.3cm}\includegraphics[scale=1.5]{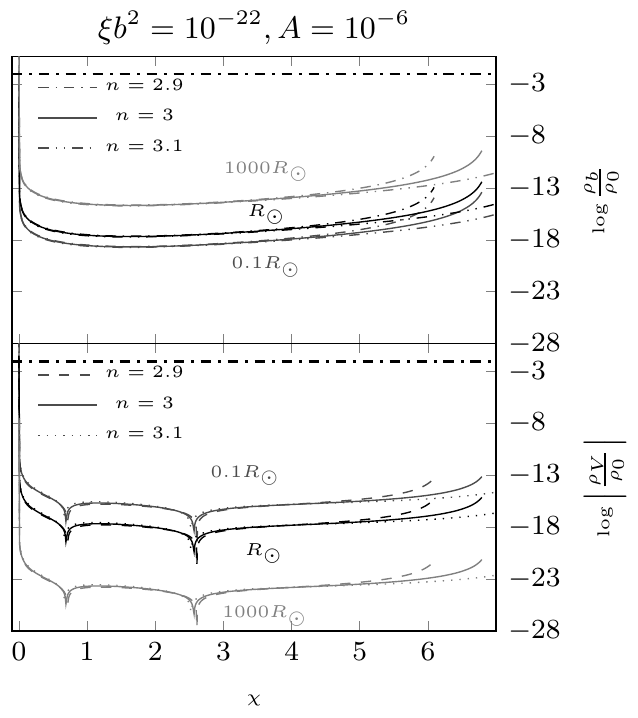}
        \end{subfigure}
       
        \begin{subfigure}[b]{0.3\textwidth}
               \hspace*{-4.5cm}\includegraphics[scale=1.5]{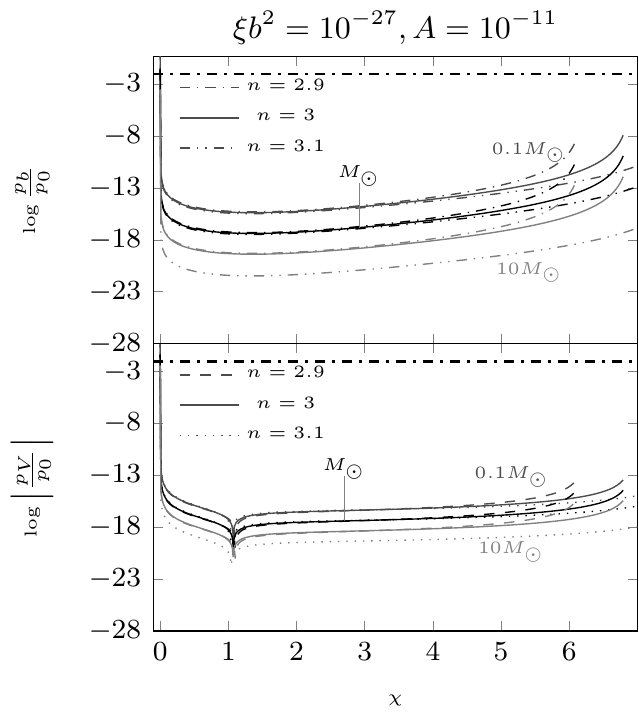}
        \end{subfigure}
        \quad
        \begin{subfigure}[b]{0.3\textwidth}
                \hspace*{-0.3cm}\includegraphics[scale=1.5]{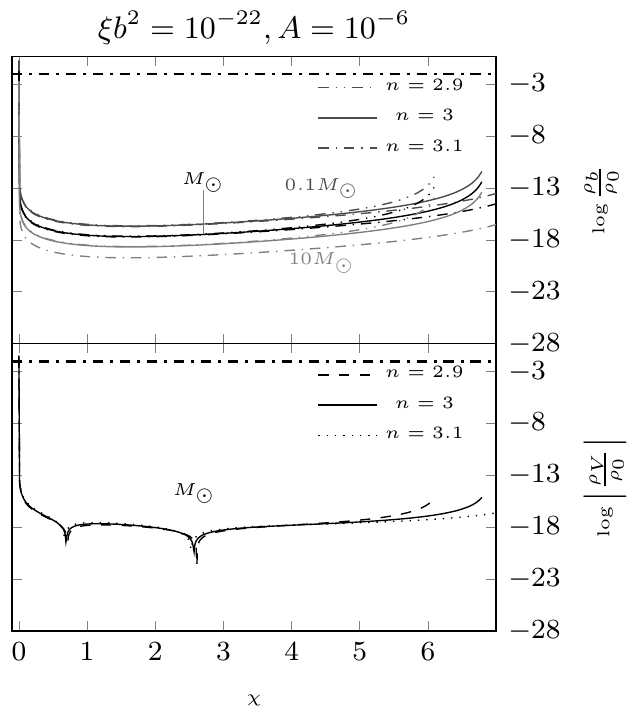}
        \end{subfigure}
        \caption{Profile of the relative perturbations $pb/p_0$, $p_V/p_0$, $\rho_b/\rho_0$ and $\rho_V/\rho_0$ induced by the bumblebee. The parameters $\xi b^2 $ and $A$ were chosen so that the maximum of the perturbations reaches the adopted $1\%$ limit.}
\label{numerical}
\end{figure*}


\subsection{Numerical analysis}

In Fig. \ref{numerical}, the profile of the contributions to Eq. (\ref{contributions}) is shown for masses and radiuses between $0.1$ and $10$ times those of the Sun. The values of the parameters $(\xi,b,A)$ are chosen so that the maximum of the relative perturbations is $1\%$, the order of magnitude of the current accuracy of the central temperature of the Sun \cite{PhysRevD.71.023521,Bertolami:2009jq,Casanellas:2011kf}.


A small variation in the polytropic index does not induce significant changes on the obtained bounds: in particular, $n$ does not impact the value of $\rho_b$, as can be seen directly in Eq. (\ref{rhob}). Increasing the radius (thus lowering $\gamma$ and $\alpha$) raises the impact of the nonvanishing VEV, while leading to a lower contribution from the potential term. A greater mass, however, leads to smaller effects on all quantities except $\rho_V$, which is rather insensitive to variations of $M$.

\begin{figure*}[!htp]
        \begin{subfigure}[b]{0.3\textwidth}
                \hspace*{-4.5cm}\includegraphics[scale=1]{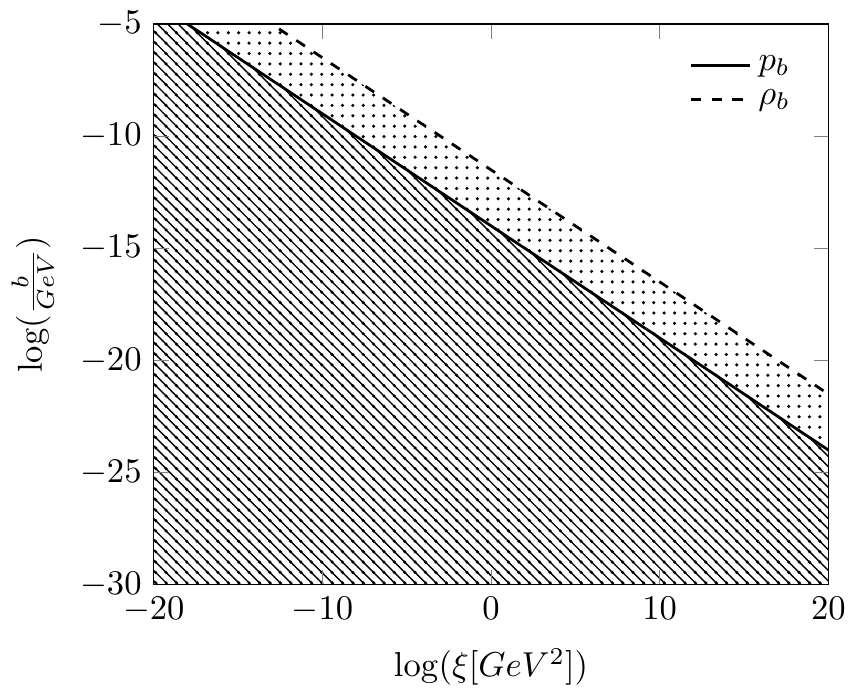}
        \end{subfigure}%
        \quad
        \begin{subfigure}[b]{0.3\textwidth}
                \hspace*{-0.3cm}\includegraphics[scale=1]{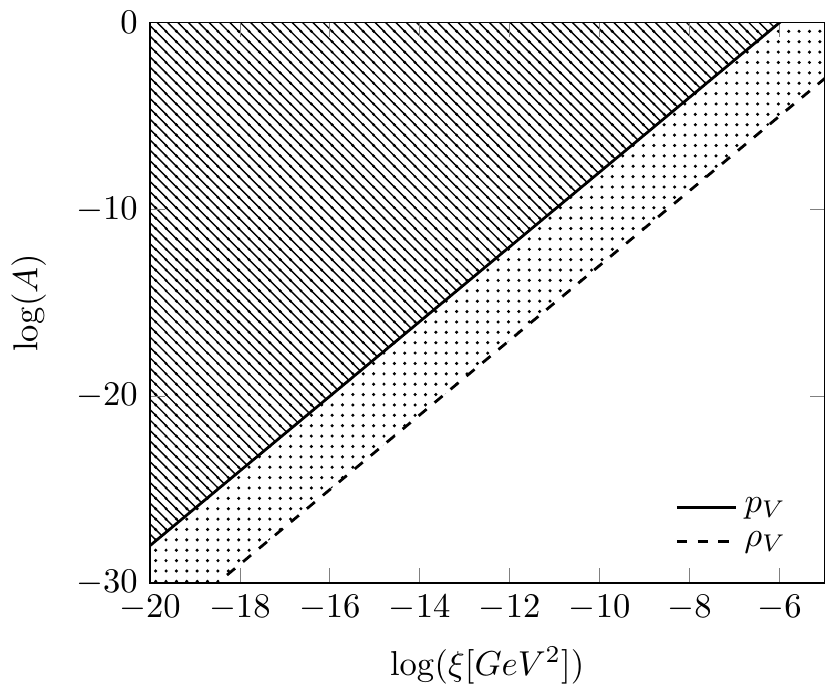}
        \end{subfigure}
\caption{Allowed region (in grey) for a relative perturbation of less than $1\%$ for $p_b$ and $\rho_b$ on the left side, and $p_V$ along with $\rho_V$ on the right side}
\label{regions}
\end{figure*}


By fixing $M=M_\odot$ and $R=R_\odot$ and finding the values for the model parameters $(\xi,b,A)$ that lead to relative perturbations of less than $1\%$, the allowed parameter space can be obtained, as depicted in Fig. \ref{regions}. Notice that the allowed values for $A$ (seen in the right panel of Fig. \ref{regions}) are bounded from below, since this quantity appears in the denominator of $p_V$ and $\rho_V$; conversely, the region allowed for $\xi b^2$ is bounded from above.

One thus obtains the bounds

\begin{equation}
\xi b^2 \lesssim 10^{-23} ~~, ~~ {\xi \over \sqrt{A} }{\rm \text{GeV}}^2 \lesssim 10^{-3} \to {\xi \over \sqrt{A} G} \lesssim 10^{34} .
\end{equation}

It is also worth noting that for the considered sets of parameters, the term $\de(\chi)$ is negligible in comparison with the other terms in both the pressure and density equations, as mentioned after Eq. (\ref{delta}): indeed, one numerically finds that $\de \lesssim 10^{-34}p_V$ (plot not shown) for the considered masses and radiuses.

\section{Discussion and Outlook}

The bumblebee field was treated as a zeroth-order perturbation on a set of stars of varying radius and mass, assuming that it follows the underlying symmetry of the problem so that it acquires only a radial component. Because the impact of the field is considered as a perturbation, an attempt was made in order to constrain the parameters of the model in such a way as to only cause a variation on the system of roughly $1\%$, following the accuracy of our present modeling of the Sun.

The obtained constraint for the value of the nonvanishing VEV of the potential driving the bumblebee field, $\xi b^2 \lesssim 10^{-23}$, is many orders of magnitude more stringent than the previously available bound $\xi b^2 \lesssim 10^{-9}$, obtained by resorting to tests of Kepler's law using the orbit of Venus \cite{Bertolami:2005bh}; by assuming that, in the presence of matter, the Bumblebee field is not relaxed at its VEV, this study has also yielded a constraint on the strength of the corresponding potential, $\xi < 10^{34} \sqrt{A} G$. Although only a quadratic potential was considered in this study, the change of the power $n$ would not dramatically change the perturbative treatment followed here.

Future refinements of this method could clearly include the use of a more accurate model for stellar structure, as well as following a more thorough numerical analysis procedure, effectively solving the (differential) modified field equations to first order in the model's parameters: this, however, should only refine the obtained bounds, with no significant change of their order of magnitude.

The application of this same methodology to the study of galaxies is also possible, in order to gain further knowledge of the constraints to the parameters of our model, as well as the possibility of describing galactic dark matter as a manifestation of the bumblebee dynamics --- following analog efforts in both scalar field \cite{PhysRevD.71.023521} and vectorial aether models \cite{Jacobson:2008aj,Barrow:2012qy,Donnelly:2010cr}. In doing so, a nonvanishing temporal component for a time-evolving Bumblebee field could also be considered, in order to provide a smooth matching at cosmological scales.

\acknowledgments \vspace{-1mm}The authors thank O. Bertolami for fruitful discussions, and the referee for his/her useful remarks. J.P. is partially supported by Funda\c c\~ao para a Ci\^encia e Tecnologia under the project PTDC/FIS/111362/2009.
\bibliography{mybib}

\end{document}